# Secure Authentication Mechanism for Cluster based Vehicular Adhoc Network (VANET): A Survey

Rabia Nasir[1], Humaira Ashraf[1], NZ Jhanjhi[2]

[1]Departement of computer science and Information Technology, International Islamic University Islamabad

[2]School of Computer Science, SCS, Taylors University, Subang Jaya, Malaysia

**Emails:** rabia.mscs1156@iiu.edu.pk, Humaira.ashraf@iiu.edu.pk , noorzaman.jhanjhi@taylors.edu.my

**Abstract:**

Vehicular Ad Hoc Networks (VANETs) play a crucial role in Intelligent Transportation Systems (ITS) by facilitating communication between vehicles and infrastructure. This communication aims to enhance road safety, improve traffic efficiency, and enhance passenger comfort. The secure and reliable exchange of information is paramount to ensure the integrity and confidentiality of data, while the authentication of vehicles and messages is essential to prevent unauthorized access and malicious activities. This survey paper presents a comprehensive analysis of existing authentication mechanisms proposed for cluster-based VANETs. The strengths, weaknesses, and suitability of these mechanisms for various scenarios are carefully examined. Additionally, the integration of secure key management techniques is discussed to enhance the overall authentication process. Cluster-based VANETs are formed by dividing the network into smaller groups or clusters, with designated cluster heads comprising one or more vehicles. Furthermore, this paper identifies gaps in the existing literature through an exploration of previous surveys. Several schemes based on different methods are critically evaluated, considering factors such as throughput, detection rate, security, packet delivery ratio, and end-to-end delay. To provide optimal solutions for authentication in cluster-based VANETs, this paper highlights AI- and ML-based routing-based schemes. These approaches leverage artificial intelligence and machine learning techniques to enhance authentication within the cluster-based VANET network. Finally, this paper explores the open research challenges that exist in the realm of authentication for cluster-based Vehicular Adhoc Networks, shedding light on areas that require further investigation and development.

## Introduction

Vehicular Adhoc Networks (VANETs) have become an integral part of modern transportation systems. These networks facilitate communication between vehicles and the infrastructure, enabling the development of advanced transportation systems. Vehicular ad hoc networks (VANETs) are a special subclass of ad hoc networks whose purpose is to enhance road safety by providing information on traffic, accidents, dangers, possible deviations, or weather information. With the emergence of autonomous vehicles and smart transportation systems, the need for secure and efficient communication networks has become a necessity. VANETs are highly dynamic networks that require a robust and reliable authentication mechanism to secure the communication channels between vehicles and the infrastructure.

The primary objective of VANETs is to ensure safety and convenience for passengers in vehicles, and therefore, optimized clustering solutions are crucial. However, the vast number of nodes and limited routers make a flat network scheme susceptible to scalability and hidden terminal issues. An effective way to address these issues is to implement a proficient clustering algorithm. As a result, there is a need to classify various VANET clustering techniques based on established criteria. Nevertheless, the clustering of vehicles faces numerous challenges such as high mobility, sparse connectivity in some regions, and security, which require well-designed solutions.**[1]**

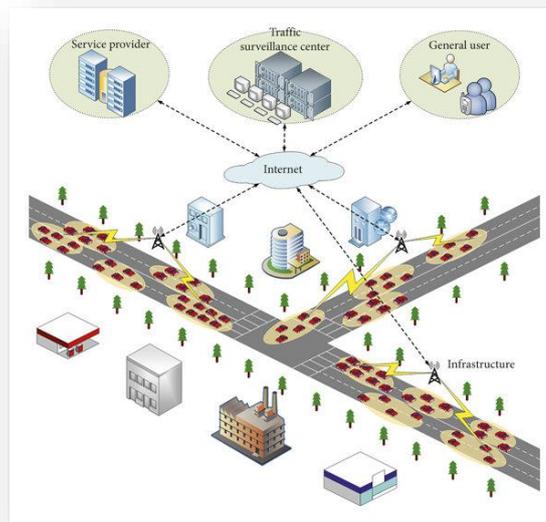

*Figure 1Cluster-based VANET architecture [2]*

In a Vehicle Ad hoc Network (VANET), vehicles share information like speed, direction, position, weather and road conditions, and obstacles with other nodes. VANETs rely on the interdependence of vehicle nodes to maintain network connection and aim to enhance road safety, manage traffic, and provide services to drivers and passengers. Due to the high number of accidents and related damages, research in VANETs is crucial. However, VANETs are vulnerable as they use wireless media, making them susceptible to attacks, where malicious nodes can send false information to deceive other vehicles, leading to adverse consequences. Therefore, security is a major challenge in VANETs that includes detecting internal and external misbehaving nodes, ensuring sender trustworthiness, and securing messages exchanged within the network.

The **[3]** focus on cryptographic techniques proposed to achieve authentication, privacy, and other security features required in VANETs like symmetric key cryptography-based schemes, public key cryptography-based schemes, identity-based cryptography schemes, pseudonym-based schemes, group and ring signature-based schemes, and blockchain-based schemes. **[4]** conducted a comprehensive analysis of recent advancements in clustering schemes for vehicular networks. The algorithms were classified based on their objective, including reliability, scalability, stability, routing overhead, and delay, as well as their purpose, including general-purpose, application-based, and technology-based clustering. The survey focused on factors such as cluster formation, maintenance, and management. **[5]** presents a comprehensive review on state-of-the-art routing protocols for UAV-aided VANETs. The protocols are categorized into seven groups in terms of their working mechanism and design principles. The shortcomings of the protocols are identified individually by critically analyzing them with regard to their advantages, disadvantages, application areas, and future improvements. The routing protocols are qualitatively compared with each other in tabular format as well on the basis of various design aspects and system parameters. **[6]** proposes eminent safety solutions to tackle the security challenges in VANETs. It covers four main areas, including attacks and security mechanisms in VANETs, a comparative analysis of security schemes based on cryptography mechanism used, trust management schemes based on discrete characteristics and intrusion detection systems, and open issues that require further investigation in the future.**[7]** review on clustering protocols in VANET. In generalized clustering protocol, cluster design aims to achieve their primary objective, i.e., the formation of a robust cluster having a long sustainable life, While in application dependent clustering protocols, cluster design aims to improve the performance of specific applications (i.e., target tracking, traffic estimation, misbehavior detection, privacy preservation, certificate revocation, etc.) on various performance metrics. **[8]** discusses the concept of vehicle clustering in Vehicular Ad Hoc Networks (VANET) to improve network performance. It provides a comprehensive taxonomy of clustering protocols in VANET based on their design objectives and explores essential research contributions of each category and objective is to improve the performance of specific applications on designated parameters such as target tracking, traffic estimation, misbehavior detection, privacy preservation, certificate revocation. conducts an extensive survey on various benchmark clustering algorithms designed by different **[9]** researchers for Cluster Head (CH) selection and reviews the routing protocols in VANET and compares various parameters for evaluating the performance of the network.

A summary of existing surveys on security and clustering techniques in Vehicular Ad hoc Networks is presented in **Table 1.** The table provides a brief overview of the motivation and contributions of these surveys, and also highlights the differences between them and the current research survey.

*Table 1 Summary of existing surveys*

| Year | Main Focus of Survey | Major Contribution | Enhancement in Our Paper |
|---|---|---|---|
| 2021 | Authentication and privacy-preserving scheme in VANETs [3] | provided the details about various security and privacy requirements in VANETs ,comprehensive survey on the existing secure authentication and privacy-preserving schemes in VANETs, comparison of the present survey with the existing surveys | Our survey presents a detailed systematic literature review of all the existing privacy and security preserving schemes in VANET |
| 2023 | Clustering in VANET[4] | clustering techniques in vehicular networks based on algorithms and applications, comparison of the present survey with the existing surveys considering the security mechanisms used, summary of cryptographic methods used in VANETs along with their key features and drawbacks | A through critical analysis of all the present protocols of clustering in vanet is provided in our survey. Also give the limitations of all existing protocols |
| 2020 | Routing Protocols for Unmanned Aerial Vehicle-Aided Vehicular Ad Hoc Networks [5] | comprehensive review of the latest routing protocols used for UAV-aided VANETs. The protocols are grouped into seven categories based on their design and working mechanism. The article also identifies the shortcomings of each protocol by critically analyzing their advantages, disadvantages, application areas, and | Our survey presents the detailed literature review along with critical analysis on all existing techniques for privacy preserving in cluster based VANET. |

| | | future improvements. The protocols are compared with each other in a tabular format based on various design aspects and system parameters, including performance, special features, optimization criteria, and techniques. The article also discusses open research issues and challenges in the field. | |
|---|---|---|---|
| 2020 | eminent safety solutions to address the security aspects for VANETs [6] | Conducted an extensive investigation on security techniques used to secure VANETs, categorizing them based on characteristics such as cryptography schemes and intrusion detection schemes. The techniques were compared based on their common properties, and open problems in enforcing security in VANETs were also presented. | Our Research presents the comprehensive critical analysis of all the techniques. |
| 2023 | State-of-the-art approach to clustering protocols in VANET [7] | classification of clustering protocols in VANET based on their design objectives, dividing them into two categories: generalized and application dependent. It explores the research contributions of each category, compares existing research on important metrics, and lists the pros and cons of existing protocols. The article aims to encourage researchers in the field of clustering in VANET by providing a comprehensive analysis of existing protocols. | Our survey presents a detailed systematic literature review of all the existing protocols of clustering |
| 2023 | Clustering in VANET [8] | conducting an extensive survey on various benchmark clustering algorithms designed by different researchers for Cluster Head (CH) selection, comparing and analyzing various parameters for evaluating the performance in a network | A through critical analysis of all the present protocols of clustering in VANET is provided pin our survey. Also give the limitations of all existing protocols |
| 2020 | provides a more thorough understanding of VANET clustering algorithms [9] | The survey provides an overview of clustering in VANETs, including definitions, cluster structure, CH selection criteria, procedural flow, and performance evaluation metrics. It presents a new taxonomy for classifying recently proposed clustering algorithms and provides a detailed description of each solution with a comparison based on relevant key parameters. The survey also highlights challenges and open research issues for each category and provides a general comparison of different clustering algorithms according to selected parameters. | Our survey also present the comprehensive critical analysis of all the systems. |

**Table 1** makes it clear that there are the number of gaps in the current research surveys. The surveys do not offer through and in-depth critical and comparative analysis, not are they presented in a structured manner. The security mechanisms in cluster based vehicular ad hoc network are not covered by the surveys. Some are on clustering techniques and some represent on overall security and privacy of VANET. . In order to contribute to the field and fill in the gaps in the current surveys, we have created this through literature review. The latest authentication mechanisms for clustered based Vehicular adhoc networks are presented in this work.

The motivation of this research was to find the secure authentication mechanism for cluster based vehicular adhoc networks (VANET). In the Systematic literature review (SLR) the process of study Firstly we select the different paper's, number of year (2020-2023) and picked 3 synonyms of each keywords of string and then for searching we used databases (Scholar, IEEE, ELSIVER, Springer) and then choose searches of first 3 pages against each strings. And then secondly discussed the inclusion criteria in which two parts one is included and the other is excluded (not included). And thirdly collect the title base and objective base paper's that are related to the research question. And also remove the duplication of papers. After that this SLR examines various methods for clustering and authentication mechanism for Vehicular Adhoc network (VANET). The techniques were analyzed in-depth, keeping specific goals in mind, and this study also provides a comprehensive critical assessment of existing methods. Based on this analysis, the SLR presents a detailed performance evaluation of all techniques, followed by a section outlining the challenges that have been identified. In conclusion, the SLR notes that numerous researchers have proposed different techniques based on various objectives.

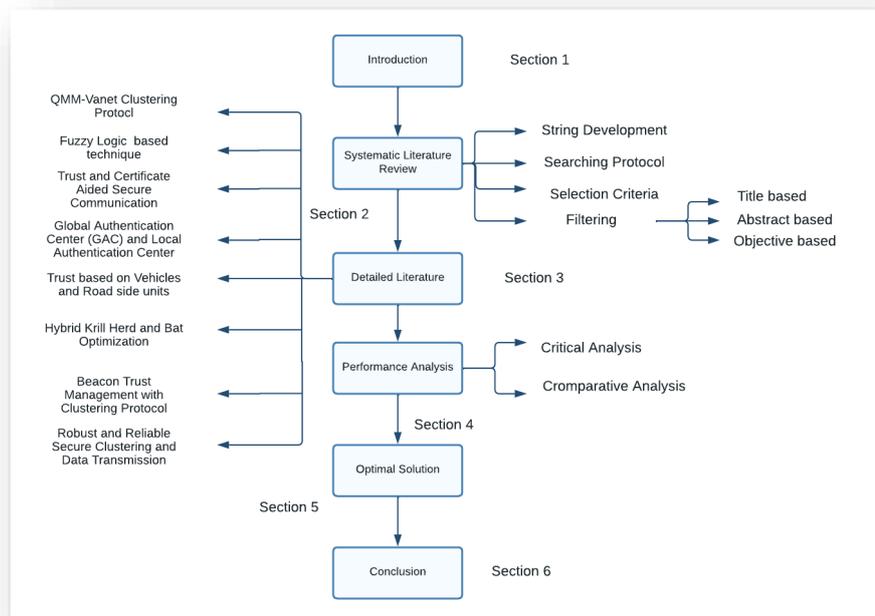

*Figure 2 Paper Organization*

The following are the key points that are discussed in this article:

- The detailed review is performed to analyze existing clustering and authentication mechanisms in VANET.
- An overview of clustering and authentication mechanisms are briefly highlighted and analyzed.
- We provide the detailed description of each existing technique.
- The drawbacks of existing Cluster based VANET and privacy preserving techniques are discussed.
- We provide the general comparison on existing authentication mechanism in cluster based VANET according to the number of parameters selected.

The organization of this paper is as follows: **Section 2** presents a systematic literature review, while **Section 3** provides an in-depth analysis of the literature. **Section 4** presents the performance analysis, which is further divided into subsections based on critical analysis, comparative analysis, and identified challenges. **Section 5** discusses the optimal solutions, and finally, Section 6 concludes the paper.

1. **Systematic Literature Review:**

Systematic Literature review (SLR) is specific approach to select secondary data. A sort of literature review known as an SLR compiles and evaluates numerous studies or papers using a methodical methodology. The aim of SLR is to offer a thorough synthesis of the literature that is available and pertinent to a research question.

The research literature discussed in this study is systematically reviewed. The systematic searches were first carried out in accordance with a searching procedure that was created. These searchers were directed by the creation of strings in accordance with the determined study query. After that, a search methodology was created in order to group all of the searches according to the search journals. The criteria for inclusion and disqualification are provided after that. Duplicates are removed using title filtering, abstract-based filtering, and objective-based filtering before publications are classified.

   **2.1 Development of String Questions:**

The strings were made using three synonyms for each keyword. Finally, applying the synonyms for each term, research papers from multiple databases are located.

Research question/String:

*Table 2 Synonyms of each word in string*

| WORDS | SYNONYM1 | SYNONYM2 | SYNONYM3 |
|---|---|---|---|
| Secure | Preserve | Surveillance | Vigilance |
| Authenticate | Authorized | Verified | Substantiate |
| Mechanism | Method | Technique | Protocol |
| Cluster | Bunch | Clump | Group |

The synonyms for each word in the string are displayed in **Table 2** Three synonyms for each term were used to create the strings. Then, research papers from various databases are found using the synonyms for each term. Total of 27 strings are created by using these synonyms for each term, and these strings are then applied to search for appropriate research papers across several databases, including IEEE, Springer, Elsevier, and Scholar.

## 1.2 SEARCHING PROTOCOL:

A procedure for searching was created, and papers published throughout a four-year period (2020, 2021, 2022, 2023) were chosen. Four databases (IEEE, Springer, Elsevier, and Scholar) and three synonyms for each phrase were also employed in the search process.

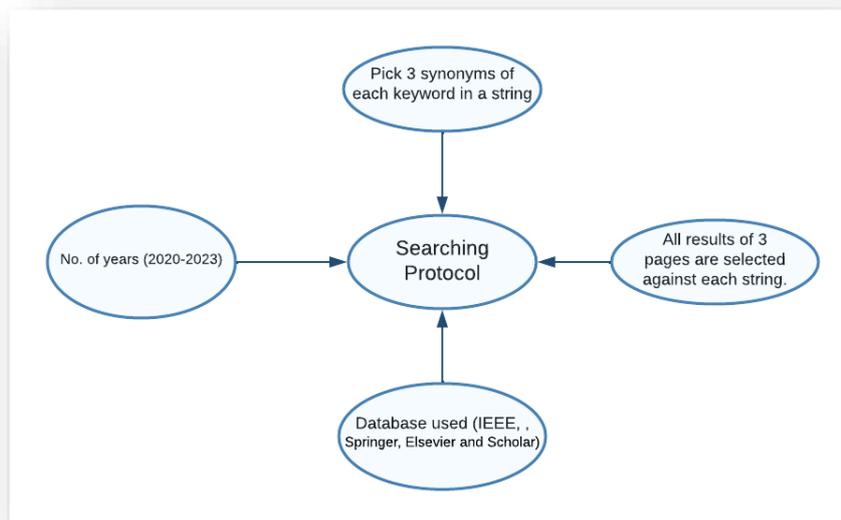

*Figure 3 Searching Protocol*

The Figure 3 shows the searching protocol. The first step is to create a string using three synonyms for each term. Second, we set the time frame from 2020 to 2023 after selecting the databases, which are IEEE, Springer, Elsevier, and Scholar. We search the publication against each string from these databases and select results from first three pages only.

### 1.3 Selection of Publication

The selection of articles for the systematic literature review was based on a number of variables, which are described in the following different sections:

### 1.3.1 Inclusion Criteria

The participation requirements contain a list of each item that research must have. Using study objectives, it was selected which piece of the research returned by the search string will be used for data collection. These are the prerequisites:

- Journals ought to provide clarification.
- Content should address a connected subject.
- Conference and journal articles ----publish research.
- There should be a paper covering the five-year period (2020 to 2023).

### 1.3.2 Exclusion Criteria

The features which would prevent research from being included are known as exclusion criteria. The exclusion criteria used to weed out articles received by the search query are listed here.

The exclusion of books, research papers that are not relevant to the topic question, newspapers, and unpublished works.

There are no additional papers that use the other languages listed.

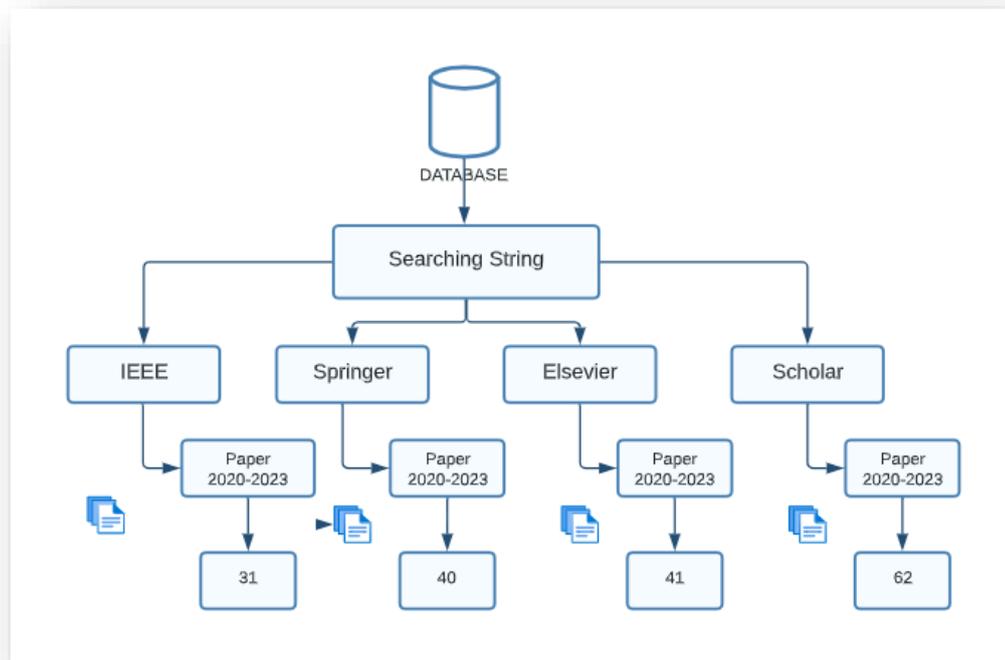

*Figure 4 Search Strategy Using Various Database*

The Figure 4 displays the number of publications we were able to track down using various databases, including IEEE, Springer, Elsevier, and Scholar. We retrieved 31 publications from IEEE, 40 papers from Springer, 41 papers from Elsevier, and 62 papers from Scholar after searching using each string in various databases.

*Table 3 Notation and definition*

| TERMS | ABBREVIATION |
|---|---|
| Qos | Quality of Service |
| MFIS | Mamdani Fuzzy interference system |
| RREQ | Route Request Message |
| RREP | Route Reply Message |
| SKC | Systematic key cryptography |
| CBP | Can Bus protocol |
| CSMA | Carrier Sense Multiple Access |
| PSO | Particle Swam Optimization |
| TBC | Trust-based clustering |
| **NS-2** | Network Simulation tool |
| **ROA** | Rainfall optimization algorithm |
| **TDMA** | Time Division Multiple Access |
| **SFIS** | Sugeno Fuzzy interference system |
| **HKH** | Hybird krill herd |
| **BOA** | Bat optimization Algorithm |
| **ZSI** | Zjdenbos similarity index |
| **SUMO** | Simulation of urban mobility |
| **FCM** | Fuzzy Cluster mechanism |
| **RSA** | Rivest-Shamir-Adleman |
| **DSA** | Digital Signature Algorithm |
| **CRL** | Certificate Revocation List |

*Table 3 List all the terms with their abbreviation that are used in this document.*

## 1.4 Filtering

Filtering is an elimination process that is used to limit our scope and reduce the number of results that are retrieved from different databases. The systematic review's goals and objectives serve as the basis for the development of exclusion criteria, which are then applied to each title and abstract in order to evaluate them. The title/abstract is disqualified if it satisfies even one of the exclusion criteria.

### 1.4.1 Title based filtering:

In title-based filtering, examine the title of each retrieved document to see if it is relevant to research question and should be included in light of the Inclusion criteria. Eliminate them if the relevance is not evident from the title.

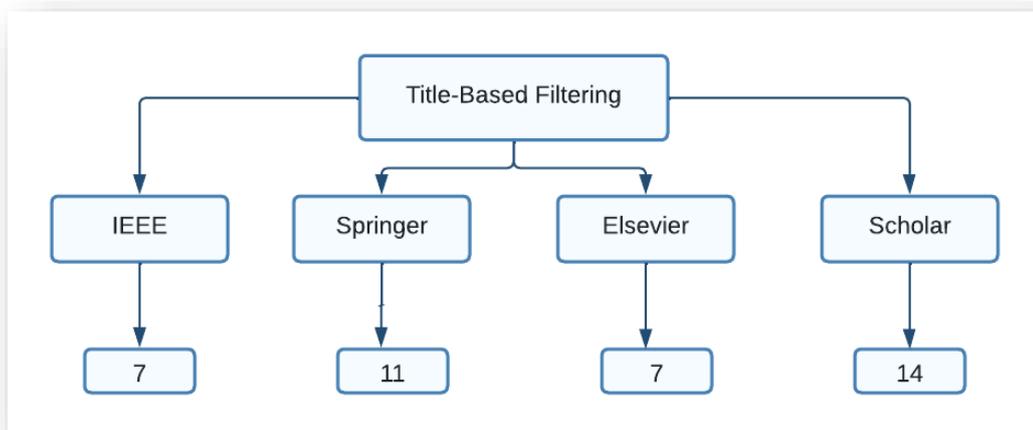

*Figure 5 Title-Based Filtering*

The Figure 5 show the title-based filtering. A total of 39 documents that are relevant to the search question/string left after title-based filtering. As shown in figure There were 4 databases; of all the papers we retrieved, we identified that 7 were relevant from the IEEE database, 11 were from Springer, 7 were from Elsevier, and 14 were from Scholar.

### 2.4.2 Abstract Based Filtering

Abstract based filtering is the second step, involves reading each retrieved publication's abstract to determine whether it is relevant to the research topic and should be included in light of the inclusion criteria. Then exclude those whose relation cannot be inferred from the abstract.

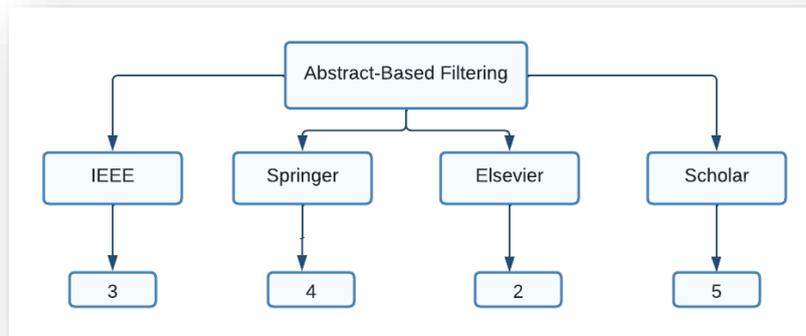

*Figure 6 Abstract-Based Filtering*

Figure 6 shows the Abstract-based filtering. A total of 23 documents that are relevant to the search question/string left after abstract-based filtering. There were 4 databases; of all the papers we retrieved, we identified that 4 were relevant from the IEEE database, 7 were from Springer, 3 were from Elsevier, and 9 were from Scholar.

### 2.4.3 Objective based filtering

After title and abstract-based filtering, objective based filtering define the aims of research papers and filter the objective of each paper. All the papers that are irrelevant are filtered according to their objective.

| Ref. | ETE | PDR | NS | NP | PLR | RE | S | IE | tH |
|---|---|---|---|---|---|---|---|---|---|
| **[10]** | ✓ | ✓ | ✓ | - |  | - | - | - |  |
| **[11]** | ✓ | ✓ | - | - | ✓ | - |  | - | ✓ |
| **[12]** |  |  | - | - |  | - | ✓ | - | ✓ |
| **[13]** | ✓ | ✓ | - | - | ✓ | ✓ |  | - | - |
| **[14]** | - |  | - | - |  |  | ✓ | ✓ | - |
| **[15]** | - | ✓ | - | - | ✓ | ✓ |  | - | - |
| **[16]** | - | - | - | ✓ | - | - | ✓ | - | ✓ |
| **[17]** | - | - | ✓ | ✓ | - | - |  | - | - |
| **[18]** | - | - | - |  | - | - | ✓ | - | - |
| **[19]** | - | - | - |  | - | - |  | - | - |
| **[20]** | ✓ | - | - | ✓ | - | - |  | - | - |
| **[21]** | - | - | - | - |  | - | ✓ | - | - |
| **[22]** | - | - | ✓ | - | - | - | - | - | ✓ |
| **[23]** | ✓ | ✓ | - | - | ✓ | - | - | - | - |

*Table 4 Objective based screening*

The objective-based filtering is displayed in **Table 4** Look for the research papers' objectives and highlight the ones listed in the table after clustering the title and abstract table-based filtering.

The objectives of papers were as follow End-to-End Delay (ETE), Packet delay route (PDR), Network Stability (NS), Network Performance (NP), Security (S), Reliability (RE), Improve Efficiency (IE), Packet Loss Rate(PLR).

### 2.4.4 Technique-based filtering

*Table 5 Technique-based filtering*

| Ref. | Technologies |
|---|---|
| [10] | Qos |
| [11] | MFIS, SKC, MAC |
| [12] | CBP, K-Mean |
| [13] | CSMA, HMAS, FCM, PSO |
| [14] | Trust based Clustering, RSUs |
| [15] | validation and certificate revocation mechanism, CRL |
| [16] | ROA |
| [17] | RSA-1024 digital signature algorithm, MAC |
| [18] | TDMA |
| [19] | SFIS |
| [20] | DSA Cryptographic Mechanism |
| [21] | BOA |
| [22] | EBTM-CP |
| [23] | Trust evaluation of vehicles, Clustering method , Cluster Head (CH) selection, Reliable route discovery |

**Table 5** Technique-based filtering show the techniques used the publications. They include protocols, mechanism, algorithms used in various papers for various purposes i.e. clustering, authentication, security etc. in order to achieve their objectives.

This *[1]* technique uses clustering based on QoS needs, distrust value factors, and mobility restrictions. The **QMM-VANET clustering protocol** is the one that has been proposed. For vehicular ad hoc networks, it is a routing protocol (VANETs).

A **fuzzy logic-based routing** technique with authentication support for vehicle ad hoc networks is the suggested method in *[2]*. Vehicle ad hoc networks (VANETs) are the technology employed, and NS2 is the simulation tool. The protocols employed in the suggested technique include hello messages, route request messages, and route reply messages.

For detecting malicious activity in the **Controller Area Network (CAN) bus** used in contemporary automobiles, the *[3]* suggests a data-driven anomaly detection system. The suggested system employs a data-driven anomaly detection method for real-time message classification as licit or illegitimate and the K-means clustering algorithm to baseline the behavior of messages moving via the CAN bus. The suggested system's other specific technologies are not mentioned in the study.

For VANET (Vehicular Ad-hoc Network), the research **[4]** suggests an unique clustering approach to boost network efficiency and identify rogue nodes for network security. The suggested clustering algorithm makes use of an advanced clustering technique to remove malicious nodes from the network and give stable cluster heads based on trust value and greatest energy. Additionally, the clustering head can be changed as needed based on a fitness function using the **PSO (Particle Swarm Optimization)** optimization technique.

No specific tool, technique, procedure, or protocol utilized for authentication is mentioned in the **[5].** The suggested technique, however, relies on trust-based clustering, which makes it easier to spot bad and compromised nodes and offers strong defense against a variety of potential threats. CRLs are difficult to keep up with. Moreover, CRLs are a poor method for rapidly disseminating important information. When a browser makes a CRL request, a CA responds by giving the browser a detailed list of all the revoked certificates it maintains.

A **Trust and Certificate Aided Secure Communication (TCASC)** System for VANET is suggested by the **[6].** Other than the suggested framework, no further specific procedures are mentioned in the paper. NS2 was the simulation programmed employed in this study. Other than the suggested scheme, no specific technology or protocol is mentioned in the study. CRL stands for Certificate Revocation List. If a node's trust value is lower than the minimum trust level in the proposed system, it is added to the CRL. The CRL data is kept up to date by the Certificate Authority (CA), which aggregates the list of cancelled identities into a single value.

Technology used in *[7]* includes the **Rainfall Optimization Algorithm (ROA)** for clustering, the Blockchain technology for secure data transmission, the Network Simulator-2 (NS-2) version 2.35 for computation, the Simulation of Urban Mobility (SUMO) for simulating vehicle mobility, and the Packet Delivery Ratio (PDR), end-to-end (ETE) delay, bandwidth, and cluster size for performance evaluation.

**Global Authentication Center (GAC) and Local Authentication Center (LAC)** are used in the *[8]* to store vehicle information and maintain a blockchain for rapid handover between internal clusters of vehicles. - IEEE 802.11 standards have modified their control packet format to fix the problems with old MAC protocols. - Processes for cluster formation, membership, cluster-head selection, merging, and departing for the transmission of safety-related and non-safety-related messages. - 5G internet with high speed for blockchain communication, and **the RSA-1024 digital signature mechanism** for sending encrypted data. - Implementation of a proof-of-concept using several virtual machines.

The *[9]* suggests a protocol for disseminating safety messages in vehicular ad-hoc networks called **Mobility-Aware Multi-hop Clustering-based MAC (MAMC-MAC)** (VANETs). The protocol creates a clustering-based multi-hop sequence using the **Time-Division Multiple Access (TDMA)** method. Using Network Simulator (NS-2) in a four-bypass motorway setting with bidirectional hops movement, the suggested protocol's performance is assessed.

For vehicle ad-hoc networks, the **[10]** suggests an improved cluster-based lifetime protocol dubbed **ECBLTR** (VANETs). The protocol incorporates both geographic and cluster-based routing protocols and evaluates the cluster head using **the Sugeno model fuzzy inference method.**

A trust-based authentication approach for clustered vehicle ad hoc networks is proposed in **[11].** (VANETs). The proposed technique is known as **TVR (Trust based on Vehicles and Road side units).** In order to ensure safe communication, the paper makes use of cryptographic techniques including digital signatures and public/private key encryption. The cluster head selection procedure is optimized in the paper by using clustering protocols to create stable clusters.

In order to overcome the problems of energy consumption and packet delay in Vehicular Ad-hoc Network, the **[12]** presents an innovative technique known as Clustered Vehicle Location protocol for **Hybrid Krill Herd and Bat Optimization (CVL-HKH-BO)** (VANET). The suggested method groups vehicles according to their locations and improves communication within each clustered. By detecting and thwarting threats based on the fitness function, the **hybrid krill herd (HKH)** and bat **optimization algorithm (BOA)** improve secure communication inside each cluster.

For vehicle ad hoc networks, the **[13]** suggests an innovative trust management protocol called Improved **Beacon Trust Management with Clustering Protocol (EBTM-CP)** (VANETs). The method entails employing **Zijdenbos similarity index (ZSI)** to determine the angle between the estimated vector and the claimed vector as well as estimating and validating the location, speed, and direction of vehicles using beacon-based trust. For effective cluster head selection and malicious node identification, EBTM-CP is used.

To overcome the security issues in vehicular ad hoc networks, the **[14]** suggests a protocol named **Robust and Reliable Secure Clustering and Data Transmission (R2SCDT/RRSCDT)** (VANETs). Any exact technique, approach, or instrument used to carry out the suggested process is not mentioned in the paper. In terms of the chosen parameters, the **TERP and TAODV protocols** have both been implemented and assessed using the NS2 tool, according to the report.

### A. Clustering mechanisms:

*Table 6 Clustering Mechanisms*

| Ref. | K-Mean | PSO | ROA | EPTM-CP |
|---|---|---|---|---|
| [12] | ✓ | - | - | - |
| [13] | - | ✓ | - | - |
| [16] | - | - | ✓ | - |
| [22] | - | - | - | ✓ |

Table 6 shows the clustering mechanisms used in different papers. The clustering method used in paper [12] is the **K-means algorithm**. It is a popular unsupervised machine learning algorithm used for clustering data points into groups based on their similarity. **PSO stands for Particle Swarm Optimization** used in paper [13] It is a popular metaheuristic optimization algorithm inspired by the social behavior of bird flocking or fish schooling. In PSO, a group of particles (also called swarm) move through a search space to find the optimal solution for a given problem. Each particle represents a potential solution, and its position and velocity are updated based on its own best position and the best position found by the swarm as a whole. The **Rainfall Optimization Algorithm (ROA) used in paper** [16] is a metaheuristic optimization algorithm inspired by the natural phenomenon of rainfall. ROA is a population-based algorithm that aims to find the optimal solution for a given problem by mimicking the behavior of rain particles. **EBTM-CP (Energy-balanced Tree-based Multi-hop Clustering Protocol)** is used in [22] . it is a clustering protocol designed for wireless sensor networks (WSNs). It is a hybrid protocol that combines a tree-based structure and multi-hop clustering to provide efficient data transmission and prolong the network lifetime.

### B. Fuzzy Mechanisms:

*Table 7 Fuzzy Logic inference system*

| Ref. | MFIS | SFIS | FCM |
|---|---|---|---|
| [11] | ✓ | - | - |
| [19] | - | ✓ | - |
| [13] | - | - | ✓ |

The Table 7 show the fuzzy interference system used in the papers. The Mamdamy fuzzy interference system (MMFIS) used in [11]**, Mamdani Fuzzy Inference Systems (MFIS)** are a type of fuzzy logic system used in decision-making processes. MFIS can be used in a wide range of applications, including control systems, pattern recognition, and decision-making processes. They are particularly useful in situations where traditional logic-based approaches are not effective due to the complexity or imprecision of the problem. **Sugeno Fuzzy Inference Systems (SFIS)** used in [19] are particularly useful in applications where the output is a function of the input variables, rather than a set of linguistic terms. They are often used in control systems, where the output is a physical parameter, such as temperature or pressure. SFIS can also be used in decision-making processes, where the output is a numerical value that represents a level of confidence or a degree of preference. **Fuzzy Cluster Mechanism (FCM) used in** [13] in which A data set is divided into N clusters, and each data point in the dataset is to some extent associated with each cluster.

### C. Cryptographic Mechanisms:

*Table 8 Cryptographic Mechanism*

| Ref. | RSA | DSA |
|---|---|---|
| [17] | ✓ | - |
| [11] | - | ✓ |

Table 8 shows the cryptographic mechanisms. **RSA(Rivest-Shamir-Adleman)** used in paper [17] and **DSA (Digital Signature Algorithm)** used in paper [11]. RSA and DSA are two widely used cryptographic mechanisms for secure communication and data transfer. RSA is a public-key cryptography algorithm that uses two keys, a public key and a private key. On the other hand, DSA is a public-key cryptography algorithm used for digital signatures. DSA is based on the mathematical concept of modular arithmetic and requires only one key, a private key. Both RSA and DSA are widely used and offer strong cryptographic security. However, RSA is generally preferred for key exchange and encryption, while DSA is preferred for digital signatures.

### D. Multiple Access Methods:

*Table 9 Multiple Access Methods*

| Ref. | CSMA | TDMA |
|---|---|---|
| [4] | ✓ | - |
| [9] | - | ✓ |

Table 9 shows the Multiple access methods. In [4] **TDMA (Time Division Multiple Access)** is used and **CSMA (Carrier Sense Multiple Access)** is used in [9] . TDMA is a method of sharing a communication channel by dividing it into time slots. Each user is allocated a specific time slot in which to transmit their data.   On the other hand, CSMA is a method of sharing a communication channel by detecting the presence of other users before transmitting data. In CSMA, a device listens to the channel before transmitting and waits until the channel is idle before transmitting.

### E. Simulation Tool:

*Table 10 Simulation tools*

| Ref. | NS-2 | NS-3 | | MATLAB |
|---|---|---|---|---|
| [10] | ✓ | - | - | - |
| [11] | ✓ | - | - | - |
| [12] | - | - | - | - |
| [13] | - | - | - | - |
| [14] | - | - | ✓ | - |
| [15] | ✓ | - | | - |
| [16] | ✓ | - | ✓ | - |
| [17] | - | - | - | - |
| [18] | ✓ | - | - | - |
| [19] | - | ✓ | - | - |
| [20] | - | - | - | ✓ |
| [21] | ✓ | - | - | - |
| [22] | ✓ | - | ✓ | - |
| [23] | ✓ | - | - | - |

The simulation tool used in research paper are shown in *Error! Reference source not found.* Table 10 Mostly papers used NS-2 simulation tool. **NS-2** (Network Simulator version 2) is an open-source event-driven network simulation tool that allows researchers to simulate and analyze the behavior of different types of wired and wireless networks. **NS-3** is also an open source discrete-event network simulator. **SUMO** (Simulation of Urban Mobility) is an open-source traffic simulation tool used to model and analyze traffic flows in urban areas. One of the papers also used **MATALB**, it is commonly used by the researchers in the field of vanets for numerical calculation and network simulation. The paper [12], [13], [17]  does not mention the specific simulator tool for proposed systems and also not mention simulation results. However paper [17]proposed method is implemented as a proof-of-concert using multiple virtual machine.

## 1.5 Detailed Literature

The [10] paper Proposed a clustering routing protocol named QMM-VANET for Vehicular Ad Hoc Networks (VANETs). The suggested protocol takes into account mobility restrictions, distrust value factors, and Quality of Service (QoS) needs. The protocol is divided into three sections: calculating the quality of service (QoS) of the vehicles and choosing a more reliable vehicle to serve as the cluster head; choosing a group of suitable neighbouring nodes to serve as gateways for retransmitting the packets; and using a gateway recovery algorithm to select a different gateway in the event that the link fails. NS-2 simulator is used to assess the proposed protocol's performance in a highway-related scenario. To show the usefulness of the suggested procedure, the simulation results are reviewed and contrasted with those from other existing protocols.

This paper [11] proposes a fuzzy logic-based routing method with authentication capability in vehicular ad hoc networks. The three phases of the suggested methodology are clustering, routing between cluster head nodes, and authentication. Four steps make up the clustering phase: choosing the cluster head nodes, joining the cluster, leaving the cluster, and supporting the cluster head nodes. Message authentication code (MAC) and symmetric key cryptography are employed during the authentication step. This study employed the simulation programme NS2 and compared the simulation findings to three additional routing protocols.

In [12], two algorithms are proposed for detecting anomalies in the data transmitted over the Controller Area Network (CAN) bus in modern vehicles. To learn the typical behaviour of messages passing on the CAN bus, a method known as the cluster-based learning algorithm is used. It gathers groups of related messages and uses them as a reference point for spotting anomalies. The data-driven anomaly detection technique is used to categorise messages as authentic or malicious in real time. It utilises the baseline established by the first algorithm to find any variations from the CAN bus' typical behaviour. Both algorithms were created to find anomalies in the CAN bus data and are based on data-driven methodologies. The experimental results demonstrate that these algorithms outperform existing approaches based on anomaly detection.

The [13] proposes a novel clustering algorithm for VANET (Vehicular Ad-hoc Network) to enhance network productivity and detect malicious nodes for network security. The suggested clustering algorithm makes use of an advanced clustering technique to remove malicious nodes from the network and give stable cluster heads based on trust value and greatest energy. Additionally, the clustering head can be changed as needed based on a fitness function using the PSO (Particle Swarm Optimization) optimisation technique. The five components that make up the proposed work are: vehicle deployment and network construction; attack development; enhanced clustering technique; PSO optimisation; and HMAC protocol for energy efficiency. The outcomes of the simulation demonstrate that the suggested clustering technique gives VANET a safe and reliable communication network.

The [14] proposed a trust-based clustering mechanism for Vehicular Ad-hoc Networks (VANETs). The proposed mechanism aims to address the challenges of preserving the stability of clusters and identifying malicious and compromised nodes. Based on a node's expertise, standing, and experience, clusters can use the trust-based clustering algorithm to identify a reliable Cluster Head (CH). Also, every node in a cluster's trust is examined to establish the backup head. A number of key issues, such as average cluster member and CH endurance, stability convergence, control overhead by speed and vehicle, throughput, and energy consumption, have been tested against the suggested mechanism. The findings demonstrate that the suggested mechanism can increase network stability and security.

The [15] propose a Trust and Certificate Aided Secure Communication (TCASC) Scheme for VANET. Vehicles are clustered in VANET according to the suggested scheme, which also calls for choosing a cluster head for each cluster and validating messages based on the opinions of other cluster members. This research uses NS2 as its simulation tool, and simulations are utilized to assess the proposed scheme's performance.

The [16] proposes a privacy-preserving data transmission architecture for cluster-based vehicular ad hoc networks (VANETs) that uses blockchain technology. The proposed approach, referred to as ROAC-B, clusters the automobiles using the rainfall optimisation algorithm (ROA) and uses blockchain-based data transmission for secure communication. clustering the network's vehicles using the rainfall optimisation algorithm (ROA). The proposed design's performance was simulated and evaluated using Simulation of Urban Mobility (SUMO) and Network Simulator-2 (NS-2) version 2.35. undertaking a series of tests to compare the proposed ROAC-B technique with other current techniques in terms of throughput, cluster size, packet delivery ratio (PDR), and end-to-end (ETE) delay. reviewing the simulation results to verify the efficacy of the suggested strategy.

The [17] proposes a multilevel blockchain-based privacy-preserving authentication protocol for Cluster-based Medium Access Control (CB-MAC) in Vehicular Ad hoc Networks (VANETs). In order to store vehicle data and maintain a blockchain for quick handover between internal clusters of cars, the suggested strategy asks for the creation of Global Authentication Centers (GAC) and Local Authentication Centers (LAC). The strategy also employs a specialised IEEE 802.11 control packet structure to alleviate the shortcomings of traditional MAC protocols. For the transmission of messages that are both safety- and non-safety-related, the plan also includes cluster formation, membership, cluster-head selection, merging, and departure activities. The proposed system transmits encrypted data using the RSA-1024 digital signature process and communicates with blockchains over fast 5G internet.

The [18] proposed the scheme the Mobility-Aware Multi-hop Clustering-based MAC (MAMC-MAC) protocol for distributing safety messages in Vehicular Ad-Hoc Networks (VANETs). The proposed scheme's methodology comprises establishing a clustering-based multi-hop sequence using the Time-Division Multiple Access (TDMA) technique. By giving all hops access to the channel and directly allocating slots, safety messages can be delivered without the need for a cluster head. reducing additional expenses while upholding impeccable honesty. evaluating the performance of the proposed protocol using Network Simulator in a four-bypass highway environment with bidirectional hops movement (NS-2). comparing the proposed protocol's performance to numerous existing protocols in terms of metrics based on QoS.

The proposed scheme in the [19] is an enhanced cluster-based lifetime protocol called ECBLTR for vehicular ad-hoc networks (VANETs). The methodology of the proposed scheme involve, Formation of clusters: The vehicles in the network are grouped into clusters based on their geographic proximity and residual energy. Selection of cluster head: The Sugeno model fuzzy inference system is used to assess the cluster head based on several input parameters, including residual energy, local distance, node degree, concentration, and distance from the base station. Routing: The proposed protocol involves both geographic and cluster-based routing protocols. The cluster head is responsible for forwarding the data packets to the base station or other cluster heads. Performance evaluation: The NS-3 simulation tool is used to evaluate the performance of the proposed protocol and compare it with other routing protocols. The performance metrics include packet delivery ratio, packet loss, average end-to-end delay, and overhead transmission. Overall, the proposed scheme aims to maximize the network's stability of routing and average throughput by using an efficient clustering and routing protocol.

The proposed scheme in the [20] is a trust-based authentication method for clustered vehicular ad hoc networks (VANETs) called TVR (Trust based on Vehicles and Road side units). The TVR approach determines each vehicle's trust degree by combining the trust between vehicles and the trust between the vehicle and Road Side Units (RSUs). The appropriate cluster head is then selected depending on the parameters of speed, direction, transmission range, number of neighbours, and trust level. The cluster heads are in charge of monitoring each vehicle and providing a secure and effective method for message delivery. The suggested method uses cryptographic technologies like public/private key encryption and digital signatures to ensure secure communication between the clusters.

The approach used in [21] proposes a unique method dubbed the Clustered Vehicle Location protocol for Hybrid Krill Herd and Bat Optimization (CVL-HKH-BO) to tackle the issues associated with energy consumption and packet delay in vehicular ad hoc networks (VANET). The suggested approach organises vehicles into clusters based on their positions and improves communication within each cluster. The hybrid krill herd and bat optimisation technique enhances secure communication inside each cluster by identifying and preventing threats based on the fitness function. After being put into practise on the Network Simulator (Ns-2) platform, the results of the recommended technique are contrasted with those of existing techniques in terms of throughput, packet loss, delay time, and data broadcasting ratio.

The [22] propose a novel trust management scheme called Enhanced Beacon Trust Management with Clustering Protocol (EBTM-CP) for vehicular ad hoc networks (VANETs). The recommended approach employs the Enhanced Beacon Trust Management with Clustering Protocol (EBTM-CP) for vehicular ad hoc networks (VANETs). The EBTM-CP approach uses beacon-based trust to estimate and validate the position, speed, and direction of vehicles. The Zijdenbos similarity index is used to calculate the angle between the estimated and claimed vectors. The suggested methodology also includes components for selecting cluster chiefs effectively and identifying rogue nodes. The simulation tool used in this paper is Network Simulator-2 (NS-2), and SUMO was used to create a real-world traffic scenario and interface with NS-2 to simulate the proposed algorithm.

The proposed scheme in [23] is called Robust and Reliable Secure Clustering and Data Transmission (R2SCDT/RRSCDT) protocol. It aims to solve the security issues in vehicular ad hoc networks by utilising vehicle trust assessments to detect malicious nodes for secure Cluster Head (CH) selection and data transfer (VANETs). The protocol involves determining the right number of clusters based on the speed and vehicle degree of each vehicle in the network, as well as safe and effective CH selection based on each vehicle's weight computation, which is done dependent on each vehicle's speed and vehicle degree. Reliable route finding for secure data transfer, where nodes are selected using a trust evaluation model and a warning is given if any node is discovered to be malicious.

*Table 11 Methodology used in proposed scheme*

| Ref. | SCHEME | METHODOLOGY |
|---|---|---|
| **[10]** | QMM-VANET | This strategy is a productive clustering algorithm that raises VANETs' QoS and security. This is accomplished by the algorithm's clustering of vehicles, resource allocation, monitoring of QoS, and detection of malicious vehicles. The simulation outcomes show how the algorithm is excellent in enhancing network performance, resource utilisation, and security. |
| **[11]** | Fuzzy logic-based routing | The three phases of the suggested technique are clustering, routing, and authentication. Selecting cluster head nodes, joining and exiting clusters, and providing support for cluster head nodes are all parts of the clustering process. The MAC algorithm and symmetric key cryptography are used for authentication. |

| [12] | Cluster-based Learning Algorithm, Data-driven Anomaly Detection Algorithm, | In order to identify malicious activity on the Controller Area Network (CAN) bus, the study suggests a data-driven anomaly detection system. The Cluster-based Learning Algorithm and the Data-driven Anomaly Detection Algorithm make up this system. Although the Data-driven Anomaly Detection Algorithm classifies messages as licit or illegitimate, the Cluster-based Learning Algorithm is used to baseline the behaviour of messages passing on the CAN bus. |
|---|---|---|
| [13] | Novel clustering algorithm | This scheme novel clustering algorithm for VANET (Vehicular Ad-hoc Network) to enhance network productivity and detect malicious nodes. It uses an advanced clustering technique, PSO optimization, CSMA data transmission, and HMAC protocol for energy efficiency. It also uses a fitness function to change the clustering head. |
| [14] | Trust-based clustering mechanism | This scheme a trust-based clustering mechanism for VANETs to address the challenges of preserving stability and identifying malicious and compromised nodes. |
| [15] | Trust and Certificate Aided Secure Communication (TCASC) | This paper proposes a Trust and Certificate Aided Secure Communication (TCASC) Scheme for VANET, which involves clustering vehicles, selecting a cluster head, and validating messages based on trust opinions. The performance of the proposed scheme is evaluated through simulations. |
| [16] | ROAC-B | The scheme is a privacy-preserving data transmission architecture for cluster-based vehicular ad hoc networks (VANETs) that uses blockchain technology. The proposed scheme clusters the vehicles using the rainfall optimization algorithm (ROA) and uses blockchain-based data transmission for secure communication. |
| [17] | A multilevel blockchain-based privacy-preserving authentication protocol | The proposed scheme involves the formation of GAC and LAC, a modified control packet format, cluster formation, membership, and cluster-head selection, and merging and leaving processes. It also uses high-speed 5G internet and RSA-1024 digital signature algorithm for encrypted information transmission. |
| [18] | Mobility-Aware Multi-hop Clustering-based MAC (MAMC-MAC) | The scheme involves the formation of GAC and LAC, a modified control packet format, cluster formation, membership, and cluster-head selection, and merging and leaving processes. It also uses high-speed 5G internet and RSA-1024 digital signature algorithm for encrypted information transmission. |
| [19] | ECBLTR | The protocol focuses on maximizing the network's stability of routing and average throughput. It uses the Sugeno model fuzzy inference system for assessing the cluster head and involves both geographic and cluster-based routing protocols. The NS-3 simulation tool is used to evaluate the performance of the proposed protocol and compare it with other routing protocols. |
| [20] | TVR (Trust based on Vehicles and Road side units). | This scheme estimates the trust degree of each vehicle by combining the trust between vehicles and the trust between the vehicle and Road Side Units (RSUs).it uses cryptographic mechanisms such as digital signatures and public/private key encryption to ensure secure communication within the clusters.<br>By increasing the range of transmission , it decrease accuracy in detecting malicious nodes. |
| [21] | Clustered Vehicle Location protocol for | The proposed technique uses a clustering approach to group vehicles based on location and optimize communication. The hybrid krill herd |

| | Hybrid Krill Herd and Bat Optimization (CVL-HKH-BO) | and bat optimization algorithm is used to detect and prevent attacks, improving secure communication. Results are compared to existing techniques in terms of throughput, packet loss, delay time, and data broadcasting ratio. |
|---|---|---|
| [22] | EBTM-CP | This scheme involves estimating and verifying the location, speed, and direction of vehicles using beacon-based trust and Zijdenbos similarity index to calculate the angle between the estimated vector and the claimed vector. The simulation tool used is Network Simulator-2 (NS-2), and SUMO was used to create a real-world traffic scenario and interface with NS-2 to simulate the proposed algorithm. |
| [23] | Robust and Reliable Secure Clustering and Data Transmission (R2SCDT/RRSCDT) protocol | This scheme designed to address the security challenges in Vehicular Ad Hoc Networks (VANETs) by using trust evaluation of vehicles to detect malicious nodes for secure Cluster Head (CH) selection and data transmission. The protocol involves discovering an optimal number of clusters, secure and optimal CH selection, and reliable route discovery for secure data transmission. The aim of the protocol is to achieve guaranteed Quality of Service (QoS) performance with minimum time and control overheads. |

**Table 11** shows the proposed schemes in paper with their methodology. The methodology explained with the techniques that are used.

## 2. Performance Analysis:

### 3.1 Critical Analysis:

The Research paper **[1]** RSUS are used to facilitate the message aggregation process by collecting the message from the vehicles in a cluster and transmitting them to the destination but Roadside Units are usually very expensive to install. Hence, authorities tend to limit their number, especially in suburbs and sparse population areas. The proposed scheme uses a QoS metric to evaluate the performance of the message aggregation process and optimize the cluster size accordingly. However, the paper does not clearly define the QoS metric used in the evaluation or provide a comparison with existing metrics. CluRMA did not consider communication between two RSUs, and privacy and security concerns have not been taken into account. In future work, communication among RSUs can be incorporated and the performance of the scheme can be analyzed while incorporating trustworthiness, privacy and security.

In [11] the proposed fuzzy logic-based routing technique with authentication capability surpasses the other three routing protocols in terms of end-to-end delay, packet collision, packet delivery rate (PDR), packet loss rate (PLR), and throughput, according to the simulation results reported in the article. The proposed approach does, however, significantly increase routing overhead. The suggested method and its three steps are thoroughly described in the paper, along with a security analysis of the approach. To provide a more thorough study, the publication might have provided more information on the clustering scheme employed in the suggested method and contrasted it with more routing protocols. Overall, the suggested solution to routing in automotive ad hoc networks with authentication support is promising. The proposed routing method uses **Mamdani fuzzy inference** to turn crisp values into fuzzy values. A defuzzifier is used to convert the fuzzy output to a crisp value. Mamdani FIS possess less flexibility in the system design[25]

There are a few limitations in [12] that should be considered. The proposed algorithms were tested on a limited dataset, and it is unclear how they would perform on a larger and more diverse dataset. The paper does not address the issue of false positives, which could lead to unnecessary alerts and potentially harm the user experience. The proposed algorithms are based on data-driven approaches, which may not be suitable for detecting sophisticated attacks that can evade anomaly detection systems. The paper does not provide a comprehensive evaluation of the proposed algorithms in terms of their computational complexity and memory requirements. The proposed algorithms are designed to detect anomalies in the CAN bus data, but they do not provide any solutions for mitigating the detected anomalies. K-mean approach is unsuitable for finding clusters that are not hyper-ellipsoids because of its sensitivity to outlier**.**[26]

The limitation [14] of is that the proposed mechanism has not been tested in a real-world scenario, and the evaluation has been done only through simulations. Another limitation is that the paper does not provide a detailed analysis of the computational overhead of the proposed mechanism. Additionally, the paper does not consider the impact of mobility on the performance of the proposed mechanism. Finally, the paper does not discuss the scalability of the proposed mechanism for large-scale VANETs. RSUs in VANETs are subject to several restrictions, because to the RSUs' limited coverage area, vehicles outside of it are unable to connect to the infrastructure. It is challenging to deploy RSUs in large quantities due to their high deployment costs. RSUs are vulnerable to attacks, and if one is successful, the network as a whole may be compromised. RSUs' restricted processing speed might lead to delays in data processing and communication.

The limitations of **[6]** include, The proposed scheme assumes that all vehicles in VANET are honest and trustworthy, which may not always be the case in real-world scenarios. The paper does not consider the impact of mobility on the proposed scheme, which can affect the performance of the scheme in dynamic environments. The proposed scheme does not address the issue of node misbehavior due to hardware or software failures. The paper does not compare the proposed scheme with other existing schemes in terms of performance and effectiveness. CRLs are challenging maintain. CRLs are also an ineffective way to instantly disseminate crucial information. A CA responds to a browser's CRL request by providing a comprehensive list of all the revoked certificates it manages. [27]

One limitation of [16] is that the proposed ROAC-B technique has not been implemented in a real-world scenario, and the simulation results may not accurately reflect the performance of the technique in a practical setting. Additionally, the paper does not consider the energy consumption of the proposed technique, which is an important factor in VANETs. The search process is weakened when inactive raindrops are removed from the Array, and it is also possible to choose the same raindrop as inactive again**.**[28]

In [17] the proposed multilevel blockchain-based privacy-preserving authentication protocol for Cluster-based Medium Access Control (CB-MAC) in Vehicular Ad hoc Networks (VANETs) appears to be a promising approach for enhancing the security and privacy of VANETs based on the information provided in the paper. The establishment of the authentication centers, car registration, and key generation operations are all fully explained in the document. The suggested architecture performs cluster creation, membership, and cluster-head selection, as well as merging and leaving processes for safety and non-safety message transmission, and it also overcomes the drawbacks of conventional MAC protocols. The implementation of the RSA-1024 digital signature algorithm and high-speed 5G internet for blockchain communication further improves the security and integrity of the system. However, additional analysis and testing might be necessary to find any The limitation of the [18] is that the proposed Mobility-Aware Multi-hop Clustering-based MAC (MAMC-MAC) protocol was evaluated only in a four-bypass highway environment with bidirectional hops movement. The protocol's performance may vary in different environments, such as urban or rural areas, with different traffic densities and patterns. Additionally, the simulation considered vehicular density as constant, which may not be the case in real-world scenarios. Therefore, further evaluation of the proposed protocol in different environments and traffic conditions is necessary to validate its effectiveness. Disadvantage using TDMA technology is that the users has a predefined time slot. When moving from one cell site to other, if all the time slots in this cell are full the user might be disconnected.[29]g scenario. The performance of the protocol may vary in a dynamic network scenario where the number of vehicles and their mobility patterns change over time. Additionally, the paper does not consider the impact of external factors such as interference and signal attenuation on the performance of the protocol. It cannot handle ambiguous data, and it cannot infer human thought. These two issues are connected to one another. A human cannot deduce knowledge or relation strings if the system's data is imprecise.**[22]**

Based on the proposed method and simulation results, the **[11]** provides a promising approach to enhance the security of vehicular ad hoc networks (VANETs). The TVR method is effective in accurately detecting malicious nodes and reducing delay and overhead. The use of clustering protocols and cryptographic mechanisms ensures stable clusters and secure communication within the clusters. However, the paper does not provide a comprehensive analysis of the limitations and challenges of the proposed method. Additionally, the simulation results are based on a specific scenario and may not be generalizable to other scenarios. Therefore, further research is needed to evaluate the proposed method in different scenarios and to address the limitations and challenges of the method. Digital signatures' main flaw is that they can only be used on a single digital document. A distinct online document, hashing technique, private key, and public key are all associated with each digital signature.[31]

In [21] proposed technique is evaluated only in a simulation environment, and its performance in a real-world scenario is not tested. The proposed technique assumes that all vehicles in the network are honest and do not participate in any malicious activity. However, in reality, some vehicles may be compromised and participate in attacks, which can affect the performance of the proposed technique. The proposed technique does not consider the impact of mobility on the network performance, and the clustering approach may not be effective in highly dynamic environments. The proposed technique requires a high computational overhead to perform the hybrid krill herd and bat optimization algorithm, which can affect the scalability of the technique. BOA has the disadvantages of being easily trapped into local minima and not being highly accurate. BOA is primarily designed for continuous optimization problems and may not be suitable for discrete optimization problems, where the search space is discrete. its performance can decrease as the problem size increases, making it less effective for solving large-scale optimization problems.**[24]**

The [22] suggested approach for vehicle ad hoc networks is based on Improved Beacon Trust Management with Clustering Protocol (EBTM-CP) (VANETs). To estimate and confirm the location, speed, and direction of vehicles, the EBTM-CP method employs beacon-based trust. The angle between the estimated and claimed vectors is calculated using the Zijdenbos similarity index. The efficient selection of cluster heads and the detection of malicious nodes are further components of the proposed methodology. The suggested EBTM-CP scheme has not been tested in a real-world environment; hence the simulation results might not correctly reflect the scheme's performance in a real-world situation. Also, the effect of network size and density on the effectiveness of the suggested system is not taken into account in the paper.

"In our survey on a secure authentication mechanism for cluster-based Vehicular Adhoc Networks (VANETs), our research draws upon foundational insights presented in [38-53]  The paper [23] proposed a  promising approach to reducing the problems caused by malevolent cars in VANETs called R2SCDT/RRSCDT protocol. To enable efficient and secure communication between vehicles, the protocol combines secure clustering with dependable data forwarding. The simulation results demonstrate that in terms of packet delivery ratio, end-to-end delay, and throughput, the proposed protocol performs better than the secure VANET communication protocols now in use. To assess the protocol's performance in real-world circumstances and with various network configurations, more analysis and testing are needed. A further drawback of the suggested approach could be the paper's lack of a thorough investigation of the protocol's security flaws and potential attack scenarios. Moreover, [54-68] serve as pillars of knowledge, contributing to the foundation upon which our exploration and analysis are grounded.

*Table 12 Summary of Critical analysis*

| Ref. | Year | Technique | Shortcoming |
|---|---|---|---|
| **[10]** | 2020 | Gateway Recovery Algorithm | The limitation of the gateway recovery algorithm proposed in the QMM-VANET protocol is that it may cause additional delay in packet delivery when a link failure occurs and a new gateway needs to be selected. This is because the algorithm requires the cluster-head to wait for a certain period of time before selecting a new gateway, which may result in increased end-to-end delay. Additionally, the algorithm may not be able to select an optimal gateway in some scenarios, which may further degrade the performance of the protocol. [26] |
| **[11]** | 2023 | Fuzzy-logic technique (MFIS) | The proposed routing method uses Mamdani fuzzy inference to turn crisp values into fuzzy values. A defuzzifier is used to convert the fuzzy output to a crisp value. Mamdani FIS possess less flexibility in the system design[25] |
| **[12]** | 2020 | k-Mean | K-mean approach is unsuitable for finding clusters that are not hyper-ellipsoids because of its sensitivity to outlier.[26] |
| **[13]** | 2020 | PSO/NSO | Particle swarm optimization (PSO) algorithm's drawbacks include its ease of entering local optimums in high-dimensional space and its slow rate of convergence during iterative processes. One of the limitations of NSO (Node Swarm Optimization) is that it may get stuck in local optima, which means that the algorithm may not be able to find the global optimal solution. This can happen when the swarm gets trapped in a region of the search space that is not the best solution, but it is still better than the other solutions in the immediate vicinity. Another limitation is that NSO may require a large number of iterations to converge to a solution, which can be time-consuming and computationally expensive. Additionally, the performance of NSO may be affected by the size of the swarm and the number of nodes in the network. |

| Ref | Year | Technique | Description |
|---|---|---|---|
| [14] | 2021 | RSUs | RSUs in VANETs are subject to several restrictions, Because to the RSUs' limited coverage area, vehicles outside of it are unable to connect to the infrastructure. It is challenging to deploy RSUs in large quantities due to their high deployment costs. RSUs are vulnerable to attacks, and if one is successful, the network as a whole may be compromised. RSUs' restricted processing speed might lead to delays in data processing and communication.[33] |
| [15] | 2020 | CRL | CRLs are challenging maintain. CRLs are also an ineffective way to instantly disseminate crucial information. A CA responds to a browser's CRL request by providing a comprehensive list of all the revoked certificates it manages.[27] |
| [16] | 2020 | ROA | The search process is weakened when inactive raindrops are removed from the Array, and it is also possible to choose the same raindrop as inactive again.[28] |
| [17] | 2021 | MAC | The design of a MAC protocol for wireless ad hoc networks has issues with node mobility, an error-prone, broadcast, shared channel, time synchronization, bandwidth efficiency, and QoS support. |
| [18] | 2021 | TDMA | The users of TDMA technology have a set time slot, which is a drawback. If all the time slots in this cell are taken when switching from one cell site to another, the user can be disconnected.[29] |
| [19] | 2023 | Fuzzy logic technique (SFIS) | Both confusing data and human mind cannot be inferred by it. These two problems are intertwined. A human cannot deduce knowledge or relation strings if the system's data is imprecise.[22] |
| [20] | 2021 | Digital Signature | Digital signatures' main flaw is that they can only be used on a single digital document. A distinct online document, hashing technique, private key, and public key are all associated with each digital signature.[31] The limitation of public/private key encryption is that it can be computationally expensive, especially for large messages. This can result in increased overhead and delay in the transmission of messages. Additionally, the security of the encryption may be compromised if the private key is stolen or if there are vulnerabilities in the encryption algorithm. The limitation of digital signature is that it does not provide confidentiality, which means that anyone can read the message. To address this limitation, encryption techniques such as public/private key encryption can be used in conjunction with digital signatures to provide both confidentiality and authentication. |
| [21] | 2021 | BOA | BOA has the drawbacks of not being very accurate and being readily stuck into local minima. Since the search space in discrete optimization issues is discrete, BOA may not be appropriate for continuous optimization problems. Its efficiency can drop as the size of the problem grows, making it less useful for handling complicated optimization issues.[24] |

| [22] | 2023 | EBTM-CP | Large-scale networks may not be a good fit for BTM-CP. EBTM-CP depends on a single CH, which is prone to failure, for each cluster. The entire cluster may be impacted if a CH fails, which could interfere with how the network functions.[13] |
| [23] | 2021 | R2SCDT/RRSCDT protocol | To assess the protocol's performance in real-world circumstances and with various network configurations, more analysis and testing are needed.[14] |

### 3.2 Comparative Analysis:

This section represents the Comparative Analysis of the performance matrices of all the schemes proposed for cluster based Vehicular Adhoc Network. All the schemes are evaluated and compared on the basics of performance matrices.

*Table 13 Comparative Analysis*

| SCHEME | ETE | PDR | NP | PLR | RE | S | IE | tH |
|---|---|---|---|---|---|---|---|---|
| QMM-VANET [10] | Low | High | - | | - | - | - | |
| Fuzzy-logic routing scheme [11] | Low | High | - | Less | - | | - | High |
| CLA [12] | - | - | Average | - | - | High | - | High |
| Secure Clustering technique for VANET. [13] | Low | High | - | Less | High | High | - | - |
| Stabtrust [14] | - | - | - | - | - | High | High | High |
| TCASC [15] | - | Average | - | Less | High | - | - | High |
| ROAC-B [16] | - | - | High | - | - | High | - | High |
| ACB-MAC [17] | Low | - | High | - | - | - | - | High |
| MAMC-MAC [18] | - | - | - | - | - | High | - | - |
| ECBLTR [19] | - | - | | - | - | - | - | - |
| TVR [20] | Low | - | High | - | - | - | - | Average |
| CVL-HKH-BO [21] | - | - | - | - | - | High | - | High |
| EBTM-CP [22] | - | - | - | - | - | - | - | High |
| | | | | | | | | |
| R2SCDT [23] | Low | High | - | Less | - | - | - | - |

The Table 13 shows the summary of the comparative analysis of the performance matrices of all the schemes proposed for Vehicular Adhoc Network. All the proposed schemes are evaluated and compared in the table based on the performance matrices.

### 3.3 Identifies Challenges:

In this section, the challenges and issues faced by all recognition schemes are presented. The limitations of these schemes are briefly outlined, highlighting areas that require further research in order to overcome the identified challenges. These areas represent open avenues for future work in recognition.

*Table 14 Identified Challenges*

| SCHEMES | CHALLENGES |
|---|---|
| **QMM-VANET [10]** | The limitation of the gateway recovery algorithm proposed in the QMM-VANET protocol is that it may cause additional delay in packet delivery when a link failure occurs and a new gateway needs to be selected. This is because the algorithm requires the cluster-head to wait for a certain period of time before selecting a new gateway, which may result in increased end-to-end delay. Additionally, the algorithm may not be able to select an optimal gateway in some scenarios, which may further degrade the performance of the protocol. |
| **Fuzzy-logic routing scheme [11]** | The proposed routing scheme has challenge that it increases the routing overhead so this can be reduce by using artificial intelligence techniques that will lower routing overhead and also increase communication security. |
| **CLA [12]** | The proposed method has some challenges that when there is large number of clusters. it is crucial for correct learning and also increase the granularity of features, it could lead to a performance reduction. On the other hand, using a few clusters could lead to performance analysis. |
| **Secure Clustering technique for VANET. [13]** | In the proposed scheme the NSO is used for clustering but there are some challenges so that it is not good to use this proposed method. One of the limitations of NSO (Node Swarm Optimization) is that it may get stuck in local optima, which means that the algorithm may not be able to find the global optimal solution. This can happen when the swarm gets trapped in a region of the search space that is not the best solution, but it is still better than the other solutions in the immediate vicinity. Another limitation is that NSO may require a large number of iterations to converge to a solution, which can be time-consuming and computationally expensive. Additionally, the performance of NSO may be affected by the size of the swarm and the number of nodes in the network. |
| **Stab trust [14]** | In the proposed method RSU acts as a central authority to formulate, coordinate and store information but the RSUs are vulnerable to attacks, and if one is successful, the network as a whole may be compromised. RSUs' restricted processing speed might lead to delays in data processing and communication. |
| **TCASC [15]** | In proposed method the increase in transmission rate **increases the Routing overhead and also decreases the delivery ratio**. Another challenge of the proposed method that when the attackers increase it decreases the |
| **ROAC-B [16]** | In this proposed method RFO is used for clustering process but there are some challenges that are, As the dataset size or dimensionality **increases, the computational complexity** of RFO grows. Efficient handling of large-scale or high-dimensional clustering problems is a challenge for RFO, as it may impact convergence speed and the quality of the clustering solution. The energy consumption of the proposed technique does not consider, which is an important factor in VANETs. These are the challenges of the proposed scheme. |
| **ACB-MAC [17]** | The proposed method has some challenges that RSA-1024 is used in it as the digital signature method, which has a security strength of 80-bits, which means at least 2^80 number of operations are required to break the keys but it is secured until the primary key is broken by the attacker. This limited |

| | key space makes it easier for attackers to perform exhaustive searches or leverage advancements in computational power to break the encryption or forge digital signatures. |
|---|---|
| **MAMC-MAC [18]** | The challenge of the proposed scheme is that it is constrained to the traffic moving in the similar direction with uniform density. when the density is on-uniform the interference is increased and bandwidth utilization is decreased. |
| **ECBLTR [19]** | The proposed method may face challenges in a dynamic network scenario where the number of vehicles and their mobility patterns change over time. Additionally, the performance of the method may be affected by external factors such as interference and signal attenuation are used in this proposed method that is computationally complex, especially when dealing with a large number of vehicles in a VANET. The complexity increases as the number of variables and rules in the fuzzy system grows. This can result in increased processing time and resource utilization, which may not be desirable in real-time VANET environments. |
| **TVR [20]** | It is difficult to work with TVR because with increase in the range of transmission, most intermediaries to search the path are the cluster head nodes, which lead to decrease accuracy in detecting malicious nodes. Also, Public/private encryption is used that not work for large messages and increased overhead and delay in the transmission of messages. Additionally, the security of the encryption may be compromised if the private key is stolen or if there are vulnerabilities in the encryption algorithm |
| **CVL-HKH-BO [21]** | It is difficult to work with the proposed method because of following challenges. The presence of malicious nodes that can compromise the security of the network and affect the performance of the proposed method. The impact of mobility on the network performance, which can affect the effectiveness of the clustering approach used in the proposed method. The high computational overhead required to perform the hybrid krill herd and bat optimization algorithm, which can affect the scalability of the proposed method. |
| **EBTM-CP [22]** | Large-scale networks may not be a good fit for BTM-CP. EBTM-CP depends on a single CH, which is prone to failure, for each cluster. The entire cluster may be impacted if a CH fails, which could interfere with how the network functions. And in these proposed methods types of attacks are not analyzed. |
| **R2SCDT [23]** | The challenge of proposed method is that there is need to assess the protocol's performance in real-world circumstances and with various network configurations, more analysis and testing are needed |

*Need to work with Decrease in packet delay:*

Proposed protocol [10] can be developed to urban scenario, utilized a security algorithm based on key distribution and also detected selfish vehicles in the network by using Swarm Intelligence methods. To address the limitations of the gateway recovery algorithm in the QMM-VANET protocol Implement predictive algorithms or machine learning models to estimate link failures and select alternative gateways proactively. Increase the trust and security in VANET by using fuzzy logic. [11]

*Need to work with Computational and Routing Overhead:*

One solution to overcome the limitations of public/private key encryption is to employ a hybrid encryption approach. This involves using a symmetric encryption algorithm to encrypt the message itself, while the public/private key encryption is used to securely exchange the symmetric encryption key. By doing so, the computational overhead associated with public/private key encryption is significantly reduced, as symmetric encryption is faster and more efficient for encrypting large amounts of data. This approach enhances security by protecting against key compromise or vulnerabilities in the encryption algorithm, while also ensuring efficient transmission of messages.

Investigating the impact of different mobility models on the performance of the proposed technique. Developing a more efficient and scalable algorithm to detect and prevent attacks in VANET. Evaluating the proposed technique in a real-world scenario to validate its effectiveness and performance. Extending the proposed technique to support multimedia data transmission in VANET. Investigating the impact of different network parameters, such as network size and density, on the performance of the proposed technique [12]. The research work can be extended further to analyses the types of specific attacks by developing the swarm intelligence method [13].

*Need to Work in detecting Malicious Nodes:*

This approach reduces delays in gateway selection and improves overall performance. The function of the proposed method [16] can be improved and limitation can be overcome using consensus algorithm. The limitation of proposed method [17] can be overcome by using Consensus protocol with the blockchain in order to collect abnormal behaviors of the vehicles from their neighbors' vehicles for behavioral analysis or reputation management. It will add some more security features, for example, removing the possibility of compromised attacks and also helping to take actions against malicious or abnormal behaviors to mitigate the presence of malicious nodes, robust security measures should be adopted, such as implementing encryption techniques, authentication protocols, and intrusion detection systems. This will enhance the security of the network and minimize the potential impact of malicious nodes on the proposed method's performance, the impact of mobility on network performance can be mitigated by employing adaptive clustering techniques that can dynamically adjust to changes in network topology. By continuously monitoring the mobility patterns and adjusting the clustering approach accordingly, the proposed method can maintain its effectiveness in the presence of mobility [21].

## 3.3.1 GAPS:

In the current literature, multiple approaches for the authentication of cluster-based VANET are proposed. However, there are still gaps present and these are:

1) Delay in Packet delivery. [10],[14], [15]
2) Increase the Routing overhead [11],[15]
3) Increase the granularity of features that lead to a performance reduction. [12],[13]
4) the computational complexity [16], [19], [21]
5) Most intermediaries to search the path are the cluster head nodes, which lead to decrease accuracy in detecting malicious nodes. [20],[22]

*Table 15 Gap Solutions*

| GAPS | SOLUTIONS |
|---|---|
| Delay in Packet delivery. [10],[14], [15] | To address the limitations of the gateway recovery that cause delay in packet delivery Implement predictive algorithms or machine |

| | | |
|---|---|---|
| | | learning models to estimate link failures and select alternative gateways proactively. [11] |
| Increase the Routing overhead [11],[15] | | The energy efficient routing protocol considers the energy constraints of individual vehicles and optimizes route selection based on energy consumption. By minimizing unnecessary communication and balancing energy usage, reduces routing overhead and extends the network lifetime.[34] Machine learning techniques to predict the optimal routes based on historical data, including traffic patterns and vehicle behavior. By making intelligent routing decisions, the proposed approach reduces routing overhead and improves network performance.[35] |
| The computational complexity [16], [19], [21] | | This gap can be reduced by using an efficient group signature scheme tailored for cluster-based VANETs. The scheme reduces computational complexity by utilizing group signature techniques, where a cluster head acts as a group manager and generates group signatures on behalf of the cluster members. This approach eliminates the need for individual vehicle signatures, significantly reducing the computational overhead associated with authentication. [36] By using machine learning-based authentication scheme that mitigates computational complexity in cluster-based VANETs. The scheme employs machine learning algorithms to learn patterns and classify incoming authentication requests, reducing the computational burden associated with traditional cryptographic verification. By selectively applying computational-intensive authentication processes based on learned patterns, this optimizes computational efficiency. [37] |
| Most intermediaries to search the path are the cluster head nodes, which lead to decrease accuracy in detecting malicious nodes. [20],[22] | | machine learning-based approach for detecting malicious nodes in cluster-based VANETs. It will utilizes historical network data, vehicle behavior, and communication patterns as input features for machine learning algorithms. By training the models to recognize patterns associated with malicious behavior, the accuracy of detecting malicious nodes is enhanced.[38] |
| Increase the granularity of features that lead to a performance reduction | | The machine learning-based feature representation technique to improve performance in cluster-based VANETs. The proposed solution employs advanced machine |

| | learning algorithms, such as deep neural networks or autoencoders, to learn compact and informative feature representations. [39] |
|---|---|

## 1. Conclusion:

This paper presents a comprehensive review of various schemes aimed at enhancing the security and authentication methods for cluster-based Vehicular Adhoc Networks (VANETs). The schemes discussed in this study encompass different approaches, including Clustering based mechanisms, Routing based schemes, Fuzzy Logic Methods, Multiple Access Methods, and Cryptographic Mechanisms. In order to evaluate and compare these schemes, an extensive critical analysis was conducted, employing a systematic literature review (SLR) methodology. The evaluation criteria encompassed various factors such as detection accuracy, complexity, packet delivery ratio, packet loss ratio, throughput, reliability, and end-to-end delay. Through this rigorous analysis, the paper identifies gaps in the existing literature, thereby highlighting areas that require further research and development in the field of cluster-based VANETs.

## REFERENCES

2. R. S. Bali, N. Kumar, and J. J. P. C. Rodrigues, "Clustering in vehicular ad hoc networks: Taxonomy, challenges and solutions," Vehicular Communications, vol. 1, no. 3, pp. 134–152, Jul. 2014, doi: 10.1016/j.vehcom.2014.05.004.
3. X. Ji, H. Yu, G. Fan, H. Sun, and L. Chen, "Efficient and Reliable Cluster-Based Data Transmission for Vehicular Ad Hoc Networks," Mobile Information Systems, vol. 2018, p. e9826782, Jul. 2018, doi: 10.1155/2018/9826782.
4. P. Mundhe, S. Verma, and S. Venkatesan, "A comprehensive survey on authentication and privacy-preserving schemes in VANETs," Computer Science Review, vol. 41, p. 100411, Aug. 2021, doi: 10.1016/j.cosrev.2021.100411.
5. "A comprehensive survey on clustering in vehicular networks: Current solutions and future challenges - ScienceDirect." https://www.sciencedirect.com/science/article/abs/pii/S1570870521002183?casa_token=XoK8oI5UptkAAAAA:-GK07s-3GVw3jkaJWccB67EYY-2DhadKWP1Raoa5eWYOOwgxnS8X_U9u6T_ZhLuKYa6jqh1pyg (accessed Apr. 28, 2023).
6. R. A. Nazib and S. Moh, "Routing Protocols for Unmanned Aerial Vehicle-Aided Vehicular Ad Hoc Networks: A Survey," IEEE Access, vol. 8, pp. 77535–77560, 2020, doi: 10.1109/ACCESS.2020.2989790.
7. A. K. Malhi, S. Batra, and H. S. Pannu, "Security of vehicular ad-hoc networks: A comprehensive survey," Computers & Security, vol. 89, p. 101664, Feb. 2020, doi: 10.1016/j.cose.2019.101664.
8. "State-of-the-art approach to clustering protocols in VANET: a survey | SpringerLink." https://link.springer.com/article/10.1007/s11276-020-02392-2 (accessed Apr. 28, 2023).
9. "Survey on Clustering in VANET Networks | IEEE Conference Publication | IEEE Xplore." https://ieeexplore.ieee.org/abstract/document/9429353?casa_token=Uva64doV1xcAAAAA:

hrF8DOXSWVjCJH4wlBU2X4OSCKTp0hJJLyIylCVgzVRwaWNXAJ4BuNAMvzmP2h3ao6wF6HlPOQ (accessed Apr. 28, 2023).
10. O. Senouci, S. Harous, and Z. Aliouat, "Survey on vehicular ad hoc networks clustering algorithms: Overview, taxonomy, challenges, and open research issues," International Journal of Communication Systems, vol. 33, no. 11, p. e4402, 2020, doi: 10.1002/dac.4402.
11. H. Fatemidokht and M. Kuchaki Rafsanjani, "QMM-VANET: An efficient clustering algorithm based on QoS and monitoring of malicious vehicles in vehicular ad hoc networks," Journal of Systems and Software, vol. 165, p. 110561, Jul. 2020, doi: 10.1016/j.jss.2020.110561.
12. M. S. Azhdari, A. Barati, and H. Barati, "A cluster-based routing method with authentication capability in Vehicular Ad hoc Networks (VANETs)," Journal of Parallel and Distributed Computing, vol. 169, pp. 1–23, Nov. 2022, doi: 10.1016/j.jpdc.2022.06.009.
13. G. D'Angelo, A. Castiglione, and F. Palmieri, "A Cluster-Based Multidimensional Approach for Detecting Attacks on Connected Vehicles," IEEE Internet of Things Journal, vol. 8, no. 16, pp. 12518–12527, Aug. 2021, doi: 10.1109/JIOT.2020.3032935.
14. A. Temurnikar, P. Verma, and J. Choudhary, "Securing Vehicular Adhoc Network against Malicious Vehicles using Advanced Clustering Technique," in 2nd International Conference on Data, Engineering and Applications (IDEA), Feb. 2020, pp. 1–9. doi: 10.1109/IDEA49133.2020.9170696.
15. K. A. Awan, I. Ud Din, A. Almogren, M. Guizani, and S. Khan, "StabTrust—A Stable and Centralized Trust-Based Clustering Mechanism for IoT Enabled Vehicular Ad-Hoc Networks," IEEE Access, vol. 8, pp. 21159–21177, 2020, doi: 10.1109/ACCESS.2020.2968948.
16. "An Improved Trust and Certificate Aided Secure Communication (TCASC) Scheme for Cluster-based VANET," Appl. Math. Inf. Sci, vol. 14, no. 1, pp. 87–95, Jan. 2020, doi: 10.18576/amis/140112.
17. G. P. Joshi, E. Perumal, K. Shankar, U. Tariq, T. Ahmad, and A. Ibrahim, "Toward Blockchain-Enabled Privacy-Preserving Data Transmission in Cluster-Based Vehicular Networks," Electronics, vol. 9, no. 9, Art. no. 9, Sep. 2020, doi: 10.3390/electronics9091358.
18. A. F. M. S. Akhter, M. Ahmed, A. F. M. S. Shah, A. Anwar, and A. Zengin, "A Secured Privacy-Preserving Multi-Level Blockchain Framework for Cluster Based VANET," Sustainability, vol. 13, no. 1, Art. no. 1, Jan. 2021, doi: 10.3390/su13010400.
19. R. Chiluveru, N. Gupta, and A. S. Teles, "Distribution of Safety Messages Using Mobility-Aware Multi-Hop Clustering in Vehicular Ad Hoc Network," Future Internet, vol. 13, no. 7, Art. no. 7, Jul. 2021, doi: 10.3390/fi13070169.
20. A. Naeem et al., "Enhanced clustering based routing protocol in vehicular ad-hoc networks," IET Electrical Systems in Transportation, vol. 13, no. 1, p. e12069, 2023, doi: 10.1049/els2.12069.
21. "A trust abased authentication method for clustered vehicular ad hoc networks | SpringerLink." https://link.springer.com/article/10.1007/s12083-020-01010-4 (accessed Mar. 22, 2023).
22. N. S. Divya, V. Bobba, and R. Vatambeti, "An Adaptive Cluster based Vehicular Routing Protocol for Secure Communication," Wireless Pers Commun, vol. 127, no. 2, pp. 1717–1736, Nov. 2022, doi: 10.1007/s11277-021-08717-4.
23. C. Gupta, L. Singh, and R. Tiwari, "Malicious Node Detection in Vehicular Ad-hoc Network (VANET) using Enhanced Beacon Trust Management with Clustering Protocol (EBTM-CP)," Wireless Pers Commun, Mar. 2023, doi: 10.1007/s11277-023-10287-6.
24. S. Hosmani and B. Mathapati, "R2SCDT: robust and reliable secure clustering and data transmission in vehicular ad hoc network using weight evaluation," J Ambient Intell Human Comput, vol. 14, no. 3, pp. 2029–2046, Mar. 2023, doi: 10.1007/s12652-021-03414-3.

25. B. Bhabani and J. Mahapatro, "CluRMA: A cluster-based RSU-enabled message aggregation scheme for vehicular ad hoc networks," Vehicular Communications, vol. 39, p. 100564, Feb. 2023, doi: 10.1016/j.vehcom.2022.100564.
26. "Comparison Between Mamdani and Sugeno Fuzzy Inference System," GeeksforGeeks, May 21, 2020. https://www.geeksforgeeks.org/comparison-between-mamdani-and-sugeno-fuzzy-inference-system/ (accessed Mar. 27, 2023).
27. "10.17148.IJARCCE.2018.7713.pdf." Accessed: Mar. 27, 2023. [Online]. Available: https://ijarcce.com/wp-content/uploads/2018/08/10.17148.IJARCCE.2018.7713.pdf
28. "What Is a Certificate Revocation List (CRL) and How Is It Used?," Security. https://www.techtarget.com/searchsecurity/definition/Certificate-Revocation-List (accessed Apr. 02, 2023).
29. S. Arulprakasam and S. Muthusamy, "Modified rainfall optimization based method for solving distributed generation placement and reconfiguration problems in distribution networks," International Journal of Numerical Modelling: Electronic Networks, Devices and Fields, vol. 35, no. 3, p. e2977, 2022, doi: 10.1002/jnm.2977.
30. "TDMA & CDMA Technologies." https://www.tutorialspoint.com/gsm/tdma_and_cdma.htm (accessed Apr. 01, 2023).
31. "Fuzzy Logic in Artificial Intelligence: Architecture, Applications, Advantages & Disadvantages," upGrad blog, Sep. 08, 2022. https://prod-eks-app-alb-1037681640.ap-south-1.elb.amazonaws.com/blog/fuzzy-login-in-artificial-intelligence/ (accessed Apr. 01, 2023).
32. "Clarifying The Differences Between Digital Signature and Electronic Signature - Trackado." https://www.trackado.com/blog/digital-signature-vs-electronic-signature/ (accessed Apr. 01, 2023).
33. L. F. Zhu, J. S. Wang, H. Y. Wang, S. S. Guo, M. W. Guo, and W. Xie, "Data Clustering Method Based on Improved Bat Algorithm With Six Convergence Factors and Local Search Operators," IEEE Access, vol. 8, pp. 80536–80560, 2020, doi: 10.1109/ACCESS.2020.2991091.
34. M. Mao, P. Yi, Z. Zhang, L. Wang, and J. Pei, "Roadside Unit Deployment Mechanism Based on Node Popularity," Mobile Information Systems, vol. 2021, p. e9980093, Jun. 2021, doi: 10.1155/2021/9980093.
35. "An energy efficient routing protocol for cluster-based wireless sensor networks using ant colony optimization | IEEE Conference Publication | IEEE Xplore." https://ieeexplore.ieee.org/abstract/document/4781748/ (accessed May 26, 2023).
36. "Mathematics | Free Full-Text | Reinforcement Learning-Based Routing Protocols in Vehicular Ad Hoc Networks for Intelligent Transport System (ITS): A Survey." https://www.mdpi.com/2227-7390/10/24/4673 (accessed May 26, 2023).
37. M. A. Saleem et al., "Deep Learning-Based Dynamic Stable Cluster Head Selection in VANET," Journal of Advanced Transportation, vol. 2021, p. e9936299, Jul. 2021, doi: 10.1155/2021/9936299.
38. G. Kaur and D. Kakkar, "Hybrid optimization enabled trust-based secure routing with deep learning-based attack detection in VANET," Ad Hoc Networks, vol. 136, p. 102961, Nov. 2022, doi: 10.1016/j.adhoc.2022.102961.
39. Sennan, S., Somula, R., Luhach, A. K., Deverajan, G. G., Alnumay, W., Jhanjhi, N. Z., ... & Sharma, P. (2021). Energy efficient optimal parent selection based routing protocol for Internet of Things using firefly optimization algorithm. Transactions on Emerging Telecommunications Technologies, 32(8), e4171.
40. Hussain, K., Hussain, S. J., Jhanjhi, N. Z., & Humayun, M. (2019, April). SYN flood attack detection based on bayes estimator (SFADBE) for MANET. In 2019 International Conference on Computer and Information Sciences (ICCIS) (pp. 1-4). IEEE.

41. Adeyemo Victor Elijah, Azween Abdullah, NZ JhanJhi, Mahadevan Supramaniam and Balogun Abdullateef O, "Ensemble and Deep-Learning Methods for Two-Class and Multi-Attack Anomaly Intrusion Detection: An Empirical Study" International Journal of Advanced Computer Science and Applications(IJACSA), 10(9), 2019. http://dx.doi.org/10.14569/IJACSA.2019.0100969
42. Lim, M., Abdullah, A., & Jhanjhi, N. Z. (2021). Performance optimization of criminal network hidden link prediction model with deep reinforcement learning. Journal of King Saud University-Computer and Information Sciences, 33(10), 1202-1210.
43. Gaur, L., Singh, G., Solanki, A., Jhanjhi, N. Z., Bhatia, U., Sharma, S., ... & Kim, W. (2021). Disposition of youth in predicting sustainable development goals using the neuro-fuzzy and random forest algorithms. Human-Centric Computing and Information Sciences, 11, NA.
44. Kumar, T., Pandey, B., Mussavi, S. H. A., & Zaman, N. (2015). CTHS based energy efficient thermal aware image ALU design on FPGA. Wireless Personal Communications, 85, 671-696.
45. Nanglia, S., Ahmad, M., Khan, F. A., & Jhanjhi, N. Z. (2022). An enhanced Predictive heterogeneous ensemble model for breast cancer prediction. Biomedical Signal Processing and Control, 72, 103279.
46. Gaur, L., Afaq, A., Solanki, A., Singh, G., Sharma, S., Jhanjhi, N. Z., ... & Le, D. N. (2021). Capitalizing on big data and revolutionary 5G technology: Extracting and visualizing ratings and reviews of global chain hotels. Computers and Electrical Engineering, 95, 107374.
47. Diwaker, C., Tomar, P., Solanki, A., Nayyar, A., Jhanjhi, N. Z., Abdullah, A., & Supramaniam, M. (2019). A new model for predicting component-based software reliability using soft computing. IEEE Access, 7, 147191-147203.
48. Hussain, S. J., Ahmed, U., Liaquat, H., Mir, S., Jhanjhi, N. Z., & Humayun, M. (2019, April). IMIAD: intelligent malware identification for android platform. In 2019 International Conference on Computer and Information Sciences (ICCIS) (pp. 1-6). IEEE.
49. Humayun, M., Alsaqer, M. S., & Jhanjhi, N. (2022). Energy optimization for smart cities using iot. Applied Artificial Intelligence, 36(1), 2037255.
50. Ghosh, G., Verma, S., Jhanjhi, N. Z., & Talib, M. N. (2020, December). Secure surveillance system using chaotic image encryption technique. In IOP conference series: materials science and engineering (Vol. 993, No. 1, p. 012062). IOP Publishing.
51. Almusaylim, Z. A., Zaman, N., & Jung, L. T. (2018, August). Proposing a data privacy aware protocol for roadside accident video reporting service using 5G in Vehicular Cloud Networks Environment. In 2018 4th International conference on computer and information sciences (ICCOINS) (pp. 1-5). IEEE.
52. Wassan, S., Chen, X., Shen, T., Waqar, M., & Jhanjhi, N. Z. (2021). Amazon product sentiment analysis using machine learning techniques. Revista Argentina de Clínica Psicológica, 30(1), 695.
53. Shahid, H., Ashraf, H., Javed, H., Humayun, M., Jhanjhi, N. Z., & AlZain, M. A. (2021). Energy optimised security against wormhole attack in iot-based wireless sensor networks. Comput. Mater. Contin, 68(2), 1967-81.
54. E. Ndashimye, N. I. Sarkar, and S. K. Ray, "A Multi-criteria based handover algorithm for vehicle-to-infrastructure communications," Computer Networks, vol. 185, no. 202152, Article ID 107652, 2020
55. Ray, S. K., Pawlikowski, K., & Sirisena, H. (2009). A fast MAC-layer handover for an IEEE 802.16 e-based WMAN. In AccessNets: Third International Conference on Access Networks, AccessNets 2008, Las Vegas, NV, USA, October 15-17, 2008. Revised Papers 3 (pp. 102-117). Springer Berlin Heidelberg.

56. Srivastava, R. K., Ray, S., Sanyal, S., & Sengupta, P. (2011). Mineralogical control on rheological inversion of a suite of deformed mafic dykes from parts of the Chottanagpur Granite Gneiss Complex of eastern India. Dyke Swarms: Keys for Geodynamic Interpretation: Keys for Geodynamic Interpretation, 263-276.
57. Ray, S. K., Sinha, R., & Ray, S. K. (2015, June). A smartphone-based post-disaster management mechanism using WiFi tethering. In 2015 IEEE 10th conference on industrial electronics and applications (ICIEA) (pp. 966-971). IEEE.
58. Chaudhuri A, Ray S (2015) Antiproliferative activity of phytochemicals present in aerial parts aqueous extract of Ampelocissus latifolia (Roxb.) planch. on apical meristem cells. Int J Pharm Bio Sci 6(2):99–108
59. Hossain, A., Ray, S. K., & Sinha, R. (2016, December). A smartphone-assisted post-disaster victim localization method. In 2016 IEEE 18th International Conference on High Performance Computing and Communications; IEEE 14th International Conference on Smart City; IEEE 2nd International Conference on Data Science and Systems (HPCC/SmartCity/DSS) (pp. 1173-1179). IEEE.
60. Airehrour, D., Gutierrez, J., & Ray, S. K. (2018). A trust-based defence scheme for mitigating blackhole and selective forwarding attacks in the RPL routing protocol. Journal of Telecommunications and the Digital Economy, 6(1), 41-49.
61. Ray, S. K., Ray, S. K., Pawlikowski, K., McInnes, A., & Sirisena, H. (2010, April). Self-tracking mobile station controls its fast handover in mobile WiMAX. In 2010 IEEE Wireless Communication and Networking Conference (pp. 1-6). IEEE.
62. Dey, K., Ray, S., Bhattacharyya, P. K., Gangopadhyay, A., Bhasin, K. K., & Verma, R. D. (1985). Salicyladehyde 4-methoxybenzoylhydrazone and diacetylbis (4-methoxybenzoylhydrazone) as ligands for tin, lead and zirconium. J. Indian Chem. Soc.;(India), 62(11).
63. Airehrour, D., Gutierrez, J., & Ray, S. K. (2017, November). A testbed implementation of a trust-aware RPL routing protocol. In 2017 27th International Telecommunication Networks and Applications Conference (ITNAC) (pp. 1-6). IEEE.
64. Ndashimye, E., Sarkar, N. I., & Ray, S. K. (2016, August). A novel network selection mechanism for vehicle-to-infrastructure communication. In 2016 IEEE 14th Intl Conf on Dependable, Autonomic and Secure Computing, 14th Intl Conf on Pervasive Intelligence and Computing, 2nd Intl Conf on Big Data Intelligence and Computing and Cyber Science and Technology Congress (DASC/PiCom/DataCom/CyberSciTech) (pp. 483-488). IEEE.
65. Ndashimye, E., Sarkar, N. I., & Ray, S. K. (2020). A network selection method for handover in vehicle-to-infrastructure communications in multi-tier networks. Wireless Networks, 26, 387-401.
66. Siddiqui, F. J., Ashraf, H., & Ullah, A. (2020). Dual server based security system for multimedia Services in Next Generation Networks. Multimedia Tools and Applications, 79, 7299-7318.
67. Shahid,H.,Ashraf,H.,Ullah,A.,Band,S.S.&Elnaffar,S.Wormholeattackmitigationstrategiesandthei rimpact onwirelesssensornetworkperformance: Aliteraturesurvey. InternationalJournalofCommunicationSystems35, e5311(2022). URLhttps://onlinelibrary.wiley.com/doi/abs/10.1002/dac.5311. https://onlinelibrary.wiley.com/doi/pdf/10.1002/dac.5311.
68. Zamir, U. B., Masood, H., Jamil, N., Bahadur, A., Munir, M., Tareen, P., ... & Ashraf, H. (2015, July). The relationship between sea surface temperature and chlorophyll-a concentration in Arabian Sea. In Biological Forum–An International Journal (Vol. 7, No. 2, pp. 825-834).